\documentclass[floatfix,aps,nofootinbib,superscriptaddress,twocolumn,10pt]{revtex4}

\pdfoutput=1
\usepackage{graphicx}
\usepackage{epstopdf}
\usepackage{amsmath}
\usepackage{amsfonts}
\usepackage{amssymb}
\usepackage{color}
\usepackage{mathrsfs}
\usepackage{multirow}
\usepackage{bm}
\usepackage[colorlinks=true, linkcolor=red, citecolor=blue]{hyperref}

\def\({\left(}
\def\){\right)}
\def\[{\left[}
\def\]{\right]}

\def\e{\begin{equation}}
\def\q{\end{equation}}
\def\m{\begin{eqnarray}}
\def\n{\end{eqnarray}}

\begin{document}

\title{Impacts of dark energy on weighing neutrinos: mass hierarchies considered}

\author{Sai Wang}
\email{physics0911@163.com}
\affiliation{Department of Physics, The Chinese University of Hong Kong, Shatin, NT, Hong Kong SAR, China}
\author{Yi-Fan Wang}
\email{yfwang@phy.cuhk.edu.hk}
\affiliation{Department of Physics, The Chinese University of Hong Kong, Shatin, NT, Hong Kong SAR, China}
\author{Dong-Mei Xia}
\email{xiadm@cqu.edu.cn}
\affiliation{College of Power Engineering, Chongqing University, Chongqing, China}
\author{Xin Zhang}
\email{zhangxin@mail.neu.edu.cn} \affiliation{Department of Physics, College of Sciences,
Northeastern University, Shenyang, China}
\affiliation{Center for High Energy Physics, Peking University, Beijing, China}


\date{\today}

\begin{abstract}

Taking into account the mass splittings between three active neutrinos, we investigate impacts of dark energy on constraining the total neutrino mass $\sum m_{\nu}$ by using recent cosmological observations. We consider two typical dark energy models, namely, the $w$CDM model and the holographic dark energy (HDE) model, which both have an additional free parameter compared with the $\Lambda$CDM model. We employ the Planck 2015 data of CMB temperature and polarization anisotropies, combined with low-redshift measurements on BAO distance scales, type Ia supernovae, Hubble constant, and Planck lensing. Compared to the $\Lambda$CDM model, our study shows that the upper limit on $\sum m_{\nu}$ becomes much looser in the $w$CDM model while much tighter in the HDE model. In the HDE model, we obtain the $95\%$ CL upper limit $\sum m_{\nu}<0.105~\textrm{eV}$ for three degenerate neutrinos. This might be the most stringent constraint on $\sum m_{\nu}$ by far and is almost on the verge of diagnosing the neutrino mass hierachies in the HDE model. However, the difference of $\chi^2$ is still not significant enough to distinguish the neutrino mass hierarchies, even though the minimal $\chi^2$ of the normal hierarchy is slightly smaller than that of the inverted hierarchy.

\end{abstract}



\maketitle

\newpage

\section{Introduction}\label{introduction}

The phenomena of neutrino oscillation indicate that there are mass splittings between three-generation active neutrinos (for a review, see \cite{Lesgourgues:2006nd}). Until recently, two independent mass squared differences have been measured by the solar and atmospheric neutrino oscillation experiments. However, it is a great challenge for the experimental particle physics to determine the absolute masses of neutrinos. In addition, the present observations have not determined whether the third neutrino is lightest or heaviest. Thus, there are two possible hierarchies for three-generation neutrinos, namely, the normal hierarchy (NH) and the inverted hierarchy (IH). Cosmological observations play an important role for studying the neutrino masses because they can provide constraints on the total neutrino mass $\sum m_\nu$ which is a key to solving the absolute masses of neutrinos by combining the results of squared mass differences. For example, Planck satellite mission \cite{Ade:2015xua} has provided tight upper bounds on the neutrino total mass.

Massive neutrinos are initially relativistic, and become non-relativistic at a transition epoch when their rest-mass begins to dominate. Their behavior has an impact on the cosmic microwave background (CMB) and the large-scale structure (LSS). Thus one can weigh the massive neutrinos through observing CMB and LSS. Massive neutrinos can affect the spectral slope of CMB anisotropies via the early-time integrated Sachs-Wolfe (ISW) effect \cite{Hou:2012xq}. They can lead to an overall modification of amplitude and location of the CMB acoustic peaks, through changing the redshift of matter-radiation equality epoch. Due to their large thermal velocity, they can suppress the clustering of matter, and then affect the gravitational lensing of the CMB photons. In addition, they can affect the angular diameter distance to the last-scattering surface, through changing the matter density in the Universe.

Recent studies have showed some tight upper limits on the total neutrino mass $\sum m_\nu$. For example, Planck satellite mission \cite{Ade:2015xua} constrained the total mass of three degenerate neutrinos in the $\Lambda$CDM model, giving $\sum m_\nu<0.72~\textrm{eV}$ ($95\%$ CL) by using Planck TT+lowP data, and $\sum m_\nu<0.49~\textrm{eV}$ ($95\%$ CL) by using Planck TT,TE,EE+lowP data. Further adding the baryon acoustic oscillation (BAO) data significantly improved the above constraints, since the BAO data can well break the acoustic scale degeneracy. The improved constraints are given, i.e., $\sum m_\nu<0.21~\textrm{eV}$ ($95\%$ CL) and $\sum m_\nu<0.17~\textrm{eV}$ ($95\%$ CL), respectively, by including the BAO measurements. If a dynamical dark energy is considered, the above constraints could be changed significantly \cite{Li:2012vn,Wang:2012uf,Zhang:2014ifa,Zhang:2015rha,Zhang:2015uhk,Allison:2015qca}. Compared to the $\Lambda$CDM model, the upper limit on $\sum m_{\nu}$ becomes much looser in the $w$CDM model while tighter in the HDE model \cite{Zhang:2015uhk}. A previous study \cite{Zhang:2015uhk} showed that the upper limit becomes $\sum m_\nu<0.113~\textrm{eV}$ ($95\%$ CL) in the HDE model.

In this study, we will focus on impacts of dynamical dark energy on constraining the neutrino total mass and distinguishing the mass hierarchies by using the latest cosmological observations. The neutrino mass has some degeneracy with the dark energy sector, since the effect of massive neutrinos on the evolution of background can be compensated by adjusting the parameters of dark energy, such as the equation-of-state parameter (EoS) $w$. Following Ref. \cite{Zhang:2015uhk}, we consider two typical dynamical dark energy models, namely, the $w$CDM model and the holographic dark energy (HDE) model, which both have an additional parameter compared to the $\Lambda$CDM model. We will show the updated constraints on the neutrino mass in both models by using the latest observations. In addition, we will compare our results with those in the $\Lambda$CDM model. To distinguish the mass hierarchies, we will further consider the neutrino mass splittings, which are obtained by the neutrino oscillation observations, in both models. A previous study \cite{Huang:2015wrx} showed that it is marginal to take these splittings into consideration in the $\Lambda$CDM model. We will show whether this previous result is still remained in this study.

The rest of this paper is arranged as follows. In Sec.~\ref{method}, we introduce the dark energy models, the method of parameter estimation, and the observational data. Our data analysis results are shown in Sec.~\ref{results}. The conclusion is given in Sec.~\ref{conclusion}.

\section{Methodology}\label{method}

We focus on two typical dark energy models, namely, the $w$CDM model and the HDE model, both have one more parameter than the $\Lambda$CDM model.

For the $w$CDM model, the parameter of EoS of dark energy is assumed to be a constant $w$.

For the HDE model \cite{Li:2004rb,Huang:2004ai}, the energy density of dark energy is given by \begin{equation}\rho_{\rm DE}=3c^2M_{\rm pl}^2R_{\rm EH}^{-2}\ ,\end{equation} where $R_{\rm EH}$ denotes the event horizon of the Universe, $M_{\rm pl}$ the reduced Planck mass, and $c$ a dimensionless parameter. The HDE model is constructed from the effective quantum field theory combined with the requirement of holographic principle of quantum gravity, and is expected to provide clues for a bottom-up exploration of a quantum theory of gravity, thus attracting extensive theoretical interests.\footnote{The cosmological constant suffers from great theoretical challenge just because we are lacking good understand for a quantum theory of gravity. The quantum field theory does not involve gravity, and thus its estimate for the vacuum energy density cannot be valid. In the current situation, the best way to try is to partly consider the effect of gravity in quantum field theory. If gravity is considered in an effective quantum field theory, then in a spatial region there cannot exist too many degrees of freedom. The holographic principle provides an effective way to exclude the extra degrees of freedom in the effective quantum field theory, which leads to the HDE model. This is why the HDE model is of great interest for many people.}
In the HDE model, the temporal evolution of EoS of dark energy is given by \cite{Li:2004rb}
\e
w(z)=-\frac{1}{3}-\frac{2}{3c}\sqrt{\Omega_{\rm DE} (z)},\label{weq}
\q
where $\Omega_{\rm DE}(z)$ is given by the solution of a differential equation. The parameter $c$ plays a crucial role in determining the evolution of dark energy in the HDE model. From Eq. (\ref{weq}), we can see that at the early times we have $w\to -1/3$, and in the infinite future we have $w\to -1/3-2/(3c)$. Thus, we find that, when $c>1$, $w$ will be always greater than $-1$, and when $c<1$, $w$ will cross $-1$ during the cosmological evolution. The value of $c$ cannot be decided by the theoretical model itself, but can only be determined by the observations.
The evolutions of $\Omega_{\rm DE}(z)$ and $H(z)$ in the HDE model incorporating massive neutrinos and dark radiation are determined by equations (2.4)--(2.7) in \cite{Zhang:2015rha}.

Since the neutrino mass effect on the cosmological background evolution can be compensated by adjusting the background parameters such as Hubble constant $H_0$ and EoS parameter $w$ (or $c$) of dark energy, the above two dynamical dark energy models would have different effects on constraints of the mass of neutrinos. For comparison, we also consider the $\Lambda$CDM model, in which the EoS parameter of the cosmological constant is $w=-1$.

We take into account the mass splittings between three active neutrinos. The observations on neutrino oscillation showed two independent mass squared differences~\cite{pdg},
\begin{eqnarray}
\label{m21}
&&\Delta m_{21}^2\equiv m_2^2-m_1^2=7.5\times10^{-5}\textrm{eV}^2\ ,\\
\label{m31}
&&|\Delta m_{31}^2|\equiv|m_3^2-m_1^2|=2.5\times10^{-3}\textrm{eV}^2\ .
\end{eqnarray}
Here the experimental uncertainties are not listed, since they are negligible when comparing to those of cosmological observations. There are two possible mass hierarchies, i.e., the normal hierarchy and the inverted one. The third-generation neutrino is heaviest for NH, while it is lightest for IH. Thus the lower limit cutoffs of total neutrino mass are $0.06~\rm{eV}$ for NH and $0.10~\rm{eV}$ for IH, respectively. For comparison, we also consider the degenerate hierarchy (DH), for which three neutrinos take the same mass.

There are six independent cosmological parameters in the base $\Lambda$CDM model, which are given by \{$\omega_b$,~$\omega_c$,~$100\theta_{\textrm{MC}}$,~$\tau,~n_s$,~$\textrm{ln}(10^{10}A_s)$\}. Here $\omega_b$ is the physical density of baryons today and $\omega_c$ is the physical density of cold dark matter today. $\theta_{\textrm{MC}}$ is the ratio between the sound horizon and the angular diameter distance at the decoupling epoch. $\tau$ is the Thomson scatter optical depth due to reionization. $n_s$ is the scalar spectrum index and $A_s$ is the amplitude of the power spectrum of primordial curvature perturbations at the pivot scale $k_p=0.05$ Mpc$^{-1}$. In addition, there is an additional free parameter $\sum m_{\nu}$ to describe the total neutrino mass. Or equivalently, one can choose the mass of the lightest neutrino as a free parameter. Thus there are seven independent parameters in total for the $\nu\Lambda$CDM model. For the $\nu$$w$CDM model, there is an extra free parameter $w$ to describe the EoS of dark energy. For the $\nu$HDE model, we also have an extra free parameter $c$ to describe the HDE. Thus there are eight independent parameters in total for both dynamical dark energy models.


To constrain cosmological parameters and neutrino mass, we employ a modified version of the publicly available Cosmological Monte Carlo (CosmoMC) sampler \cite{Lewis:2002ah} to estimate the parameter space of cosmological models. For NH model, the neutrino mass spectrum is written as
\e
(m_1,m_2,m_3) = (m_1,\sqrt{m_1^2+\Delta m_{21}^2},\sqrt{m_1^2+|\Delta m_{31}^2|}),
\q
in terms of a free parameter $m_1$. While for IH the neutrino mass spectrum is expressed in terms of $m_3$,
\e
(m_1,m_2,m_3) = (\sqrt{m_3^2+|\Delta m_{31}^2|},\sqrt{m_3^2+|\Delta m_{31}^2|+\Delta m_{21}^2},m_3).
\q
For DH the default setting of neutrino mass spectrum is invoked
\e
m_1=m_2=m_3=m,
\q
where $m$ is a free parameter. It should also be noted that the input lower bound of $\sum m_\nu$ is $0.06$ eV, 0.10 eV, and $0$ for NH, IH, and DH, respectively. The priors of all the base parameters are set to be uniform in this study.


In our study, we use Planck 2015 data release of CMB temperature and polarization anisotropies (denoted by Planck TT,TE,EE+lowP) \cite{Aghanim:2015xee}, combined with other low-redshift measurements. As an alternative, one can use the distance priors to summarize the Planck CMB data \cite{Huang:2015vpa}. The BAO distance scales are considered to break the geometric degeneracy. We use the LOWZ and CMASS samples of BOSS DR12 \citep{Cuesta:2015mqa}, as well as the 6dFGS \citep{Beutler:2011hx} and the SDSS MGS \citep{Ross:2014qpa}. Thus our basic data combination is denoted by Planck TT,TE,EE+lowP+BAO.

To further constrain parameters of dark energy, we consider more low-redshift measurements, including the type Ia supernovae (SN), Hubble constant ($H_0$), and Planck lensing. We use the ``joint light-curve analysis'' (JLA) compilation of the SN data \cite{Betoule:2014frx}. We use the local measurement of $H_0$, i.e., $H_0=73.02\pm1.79\rm{~km~s^{-1}Mpc^{-1}}$ at $1\sigma$ confidence level (CL) \cite{Riess:2016jrr}. The uncertainty of this measurement has been reduced to $2.4\%$. In addition, we also use Planck lensing data \cite{Ade:2015zua}, which provide additional information at low redshift. Thus our other data combination is denoted by Planck TT,TE,EE+lowP+BAO+JLA+$H_0$+Lensing.

\section{Results}\label{results}
\begin{table*}[!htp]
\centering
\renewcommand{\arraystretch}{1.5}
\scalebox{1}[1]{%
\begin{tabular}{|c|c c c|c c c|}
\hline
&\multicolumn{3}{c|}{$\textit{Planck}$ TT,TE,EE+lowP+BAO}  &  \multicolumn{3}{c|}{$\textit{Planck}$ TT,TE,EE+lowP+BAO+JLA+$H_0$+Lensing}     \\
\cline{2-7}
&$\nu_{\textrm{NH}}\Lambda \textrm{CDM}$&$\nu_{\textrm{IH}}\Lambda \textrm{CDM}$&$\nu_{\textrm{DH}}\Lambda \textrm{CDM}$&$\nu_{\textrm{NH}}\Lambda \textrm{CDM}$&$\nu_{\textrm{IH}}\Lambda \textrm{CDM}$&$\nu_{\textrm{DH}}\Lambda \textrm{CDM}$\\
\hline
$\Omega_bh^2$&$		0.02230\pm0.00014$&	$0.02232\pm0.00014$&	$0.02229\pm0.00014$&	 $0.02238\pm0.00014$ &	$0.02239\pm0.00014$&	$0.02236\pm0.00014$   \\
$\Omega_ch^2$&$		0.1188\pm0.0011$&		$0.1186\pm0.0011$&	$0.1191\pm0.0011$&	 $0.1177\pm0.0011$&	$0.1175\pm0.0010$&	$0.1180\pm0.0011$ \\
$100\theta_{\emph{\rm MC}}$&$1.04086\pm0.00030$& $1.04087\pm0.00030$&	 $1.04085\pm0.00030$&	 $1.04103\pm0.00029$ &	$1.04104\pm0.00030$&	 $1.04102\pm0.00029$   \\
$\tau$&$				0.085\pm0.017$&		$0.088\pm0.01$&		$0.082\pm0.017$&		 $0.075\pm0.013$&		$0.080\pm0.013$&		$0.070\pm0.014$\\
${\textrm{ln}}(10^{10}A_s)$&$3.103\pm0.032$&	$3.108\pm0.032$&		$3.097\pm0.033$&		 $3.080\pm0.024$&		$3.088\pm0.024$&		$3.070\pm0.025$  \\
$n_s$&$				0.9670\pm0.0042$&		$0.9677\pm0.0042$&	$0.9663\pm0.0042$&	 $0.9695\pm0.0041$&	$0.9700\pm0.0041$&	$0.9686\pm0.0041$ \\
$\sum m_\nu$ &$		<0.179~\textrm{eV}$&	$<0.203~\textrm{eV}$&	$<0.157~\textrm{eV}$&	 $<0.158~\textrm{eV}$  &	$<0.186~\textrm{eV}$&	$<0.134~\textrm{eV}$   \\
\hline
$H_0$&				$67.42\pm0.55$&		$67.19\pm0.53$&		$67.65\pm0.58$&		 $67.99\pm0.51$&		$67.75\pm0.49$&		$68.23\pm0.55$ \\
\hline
$\chi^2_{\rm min}$&		$12947.0$&			$12948.7$&			$12946.4$&			 $13663.5$&			 $13665.3$&			$13661.9$ \\
\hline
\end{tabular}}
\caption{Constraints on six independent cosmological parameters ($68\%$ CL) and neutrino mass ($95\%$ CL) in the $\Lambda$CDM model with three active neutrinos of NH, IH, and DH, respectively.}
\label{tab:lcdm}
\end{table*}

\begin{table*}[!htp]
\centering
\renewcommand{\arraystretch}{1.5}
\scalebox{1}[1]{%
\begin{tabular}{|c|c c c|c c c|}
\hline
&\multicolumn{3}{c|}{$\textit{Planck}$ TT,TE,EE+lowP+BAO}  &  \multicolumn{3}{c|}{$\textit{Planck}$ TT,TE,EE+lowP+BAO+JLA+$H_0$+Lensing}     \\
\cline{2-7}
&$\nu_{\textrm{NH}}w \textrm{CDM}$&$\nu_{\textrm{IH}}w \textrm{CDM}$&$\nu_{\textrm{DH}}w \textrm{CDM}$&$\nu_{\textrm{NH}}w \textrm{CDM}$&$\nu_{\textrm{IH}}w \textrm{CDM}$&$\nu_{\textrm{DH}}w \textrm{CDM}$\\
\hline
$\Omega_bh^2$&		$0.02224\pm0.00015$&	$0.02224\pm0.00015$&	$0.02225\pm0.00015$&	 $0.02228\pm0.00015$ &	$0.02227\pm0.00015$&	$0.02227\pm0.00015$   \\
$\Omega_ch^2$&		$0.1197\pm0.0014$&	$0.1197\pm0.0014$&	$0.1197\pm0.0014$&	 $0.1190\pm0.0012$&	$0.1189\pm0.0012$&	$0.1191\pm0.0012$ \\
$100\theta_{\emph{\rm MC}}$&$1.04075\pm0.00032$&	$1.04074\pm0.00032$&	 $1.04078\pm0.00031$&	$1.04087\pm0.00031$ &	$1.04086\pm0.00031$&	 $1.04087\pm0.00030$   \\
$\tau$&				$0.082\pm0.017$&		$0.082\pm0.017$&		$0.080\pm0.017$&		 $0.068\pm0.015$&		$0.070\pm0.015$&		$0.066\pm0.016$\\
${\textrm{ln}}(10^{10}A_s)$&$3.098\pm0.033$&	$3.099\pm0.032$&		$3.095\pm0.033$&		 $3.068\pm0.028$&		$3.073\pm0.027$&		$3.063\pm0.029$  \\
$n_s$&				$0.9648\pm0.0046$&	$0.9647\pm0.0046$&	$0.9648\pm0.0045$&	 $0.9662\pm0.0044$&	$0.9661\pm0.0044$&	$0.9659\pm0.0044$ \\
$w$&				$-1.094\pm0.088$&		$-1.110\pm0.087$&		$-1.073\pm0.086$&		 $-1.087\pm0.048$&		$-1.096\pm0.048$&		$-1.080\pm0.050$ \\
$\sum m_\nu$ &		$<0.287~\textrm{eV}$&	$<0.302~\textrm{eV}$&	$<0.273~\textrm{eV}$&	 $<0.274~\textrm{eV}$  &	$<0.287~\textrm{eV}$&	$<0.268~\textrm{eV}$   \\
\hline
$H_0$&				$69.40\pm1.91$&		$69.59\pm1.93$&		$69.16\pm1.84$&	 $69.60\pm1.00$&		$69.61\pm0.99$&			$69.58\pm0.99$ \\
\hline
$\chi^2_{\rm min}$&		$12947.0$&			$12947.8$&			$12946.5$&			 $13660.9$&			 $13663.0$&			$13661.2$ \\
\hline
\end{tabular}}
\caption{Constraints on seven independent cosmological parameters ($68\%$ CL) and neutrino mass ($95\%$ CL) in the wCDM model with three active neutrinos of NH, IH, and DH, respectively.}
\label{tab:wcdm}
\end{table*}

\begin{table*}[!htp]
\centering
\renewcommand{\arraystretch}{1.5}
\scalebox{1}[1]{%
\begin{tabular}{|c|c c c|c c c|}
\hline
&\multicolumn{3}{c|}{$\textit{Planck}$ TT,TE,EE+lowP+BAO}  &  \multicolumn{3}{c|}{$\textit{Planck}$ TT,TE,EE+lowP+BAO+JLA+$H_0$+Lensing}     \\
\cline{2-7}
&$\nu_{\textrm{NH}}\textrm{HDE}$&$\nu_{\textrm{IH}} \textrm{HDE}$&$\nu_{\textrm{DH}}\textrm{HDE}$&$\nu_{\textrm{NH}} \textrm{HDE}$&$\nu_{\textrm{IH}}\textrm{HDE}$&$\nu_{\textrm{DH}}\textrm{HDE}$\\
\hline
$\Omega_bh^2$&		$0.02228\pm0.00015$&	$0.02228\pm0.00015$&	$0.02228\pm0.00015$&	 $0.02238\pm0.00015$ &	$0.02238\pm0.00015$&	$0.02237\pm0.00015$   \\
$\Omega_ch^2$&		$0.1191\pm0.0013$&	$0.1190\pm0.0013$&	$0.1191\pm0.0013$&	 $0.1176\pm0.0012$&	$0.1175\pm0.0012$&	$0.1176\pm0.0012$ \\
$100\theta_{\emph{\rm MC}}$&$1.04084\pm0.00031$&	$1.04083\pm0.00031$&	 $1.04085\pm0.00031$&	$1.04105\pm0.00030$ &	$1.04103\pm0.00031$&	 $1.04107\pm0.00030$   \\
$\tau$&				$0.087\pm0.017$&		$0.089\pm0.017$&		$0.086\pm0.017$&		 $0.087\pm0.014$&		$0.090\pm0.013$&		$0.084\pm0.013$\\
${\textrm{ln}}(10^{10}A_s)$&$3.108\pm0.033$&	$3.112\pm0.033$&		$3.105\pm0.033$&		 $3.103\pm0.025$&		$3.110\pm0.025$&		$3.097\pm0.025$  \\
$n_s$&				$0.9663\pm0.0045$&	$0.9665\pm0.0046$&	$0.9663\pm0.0046$&	 $0.9700\pm0.0044$&	$0.9703\pm0.0045$&	$0.9698\pm0.0044$ \\
$c$&					$0.501\pm0.049$&		$0.488\pm0.047$&		$0.519\pm0.051$&		 $0.590\pm0.030$&		$0.581\pm0.029$&		$0.603\pm0.030$ \\
$\sum m_\nu$ &		$<0.196~\textrm{eV}$&	$<0.223~\textrm{eV}$&	$<0.163~\textrm{eV}$&	 $<0.145~\textrm{eV}$  &	$<0.173~\textrm{eV}$&	$<0.105~\textrm{eV}$   \\
\hline
$H_0$&				$73.23\pm2.46$&		$73.57\pm2.52$&		$72.79\pm2.38$&		 $69.66\pm0.98$&		$69.64\pm0.98$&	$69.67\pm0.97$ \\
\hline
$\chi^2_{\rm min}$&		$12954.2$&			$12955.7$&			$12953.5$&			 $13673.3$&			 $13677.4$&			$13672.5$ \\
\hline
\end{tabular}}
\caption{Constraints on seven independent cosmological parameters ($68\%$ CL) and neutrino mass ($95\%$ CL) in the HDE model with three active neutrinos of NH, IH, and DH, respectively.}
\label{tab:hde}
\end{table*}

Our constraints on cosmological parameters are listed in Tables~\ref{tab:lcdm}--\ref{tab:hde} for the $\Lambda$CDM model, the $w$CDM model, and the HDE model, respectively. For each particular dark energy model, we further consider three neutrino mass hierarchy models which are denoted by $\nu_\textrm{NH}$, $\nu_\textrm{IH}$, and $\nu_\textrm{DH}$, respectively. The best-fit results of the $\Lambda$CDM model are listed here for comparison with the other two dark energy models. In these tables, we show our best-fit results with the $68\%$ CL uncertainty for the cosmological parameters, but we make the $95\%$ CL upper limits for the neutrino mass $\sum m_\nu$. In addition, the best-fit $\chi^2$ are also listed here, as well as the derived parameter $H_0$.

By using the Planck TT,TE,EE+lowP+BAO data, the upper limits ($95\%$ CL) on the neutrino total mass in the $\Lambda \textrm{CDM}$ model are obtained as
$\{\sum m_{\nu_{\textrm{NH}}},~\sum m_{\nu_{\textrm{IH}}},~\sum m_{\nu_{\textrm{DH}}}\}_{\Lambda \textrm{CDM}}<\{0.179,~0.203,~0.157\}~\textrm{eV}$,
while in the $w$CDM model the results are
$\{\sum m_{\nu_{\textrm{NH}}},~\sum m_{\nu_{\textrm{IH}}},~\sum m_{\nu_{\textrm{DH}}}\}_{w \textrm{CDM}}<\{0.287,~0.302,~0.273\}~\textrm{eV}$
and in the $\textrm{HDE}$ model
$\{\sum m_{\nu_{\textrm{NH}}},~\sum m_{\nu_{\textrm{IH}}},~\sum m_{\nu_{\textrm{DH}}}\}_{\textrm{HDE}}<\{0.196,~0.223,~0.163\}~\textrm{eV}$, respectively.
In the $w$CDM model, the upper limit of neutrino total mass is replaced by a much looser value compared with the $\Lambda \textrm{CDM}$ model, while in the $\textrm{HDE}$ model the upper limit is slightly looser.

Including the low-redshift observations helps further break the parameter degeneracy. In the case of Planck TT,TE,EE+lowP+BAO+JLA+$H_0$+Lensing fit, it shows that tighter bounds are obtained for the $\sum m_\nu$ estimation, compared to the identical models without adding the low-redshift data. Compared with the $\Lambda$CDM model, the upper limit on $\sum m_{\nu}$ becomes much looser in the $w$CDM model while much tighter in the HDE model, consistent with the results in \cite{Zhang:2015uhk}. Our best-fit results show that the neutrino mass splittings can indeed affect the upper limits on the neutrino total mass for the HDE model. But for the $\Lambda$CDM model and the $w$CDM model, the mass splitting only marginally affects the constraints on neutrino total mass, which is consistent with \cite{Huang:2015wrx}. Particularly, with $\textrm{HDE}$ model and  degeneracy neutrino mass hierarchy considered, $\sum m_\nu < 0.105 ~\textrm{eV}$ is obtained. This value may be the most strict constraint on $\sum m_\nu$ by far and is already on the verge of the lower bound given by  the neutrino IH model. This result implies that the neutrino DH model is not convincing in the HDE model and the mass splitting effect has to be considered.
In addition, Planck lensing data can indeed improve the measurement of $\tau$. After adding Planck lensing, the values of $\tau$ are significantly lowered for $\Lambda$CDM and $w$CDM, but still remained unchanged for HDE.

The values of  $\chi^2_\textrm{min}$ of different models in the fit are also listed. For the $w$CDM model, the values of $\chi^2_\textrm{min}$ are almost the same as those of the $\Lambda \textrm{CDM}$ model without considering low-redshift data, and are slightly smaller than those after considering low-redshift data, at the price of having one more parameter. For the HDE model, the $\chi^2_\textrm{min}$  values are much larger than those of $\Lambda \textrm{CDM}$ model. The reason is that the HDE model does not fit the BAO data at $z_{\textrm{eff}}=0.57$ and the JLA data well in the global best fitting. For different neutrino mass hierarchy models, even though the normal hierarchy's $\chi^2_\textrm{min}$ is slightly smaller than that of the inverted hierarchy, the difference $\Delta \chi^2\equiv\chi^2_{\rm IH,min}-\chi^2_{\rm NH,min}$ is not significant enough to distinguish the hierarchies.

\begin{figure}[htbp]
\centering
\includegraphics[width=3.4in]{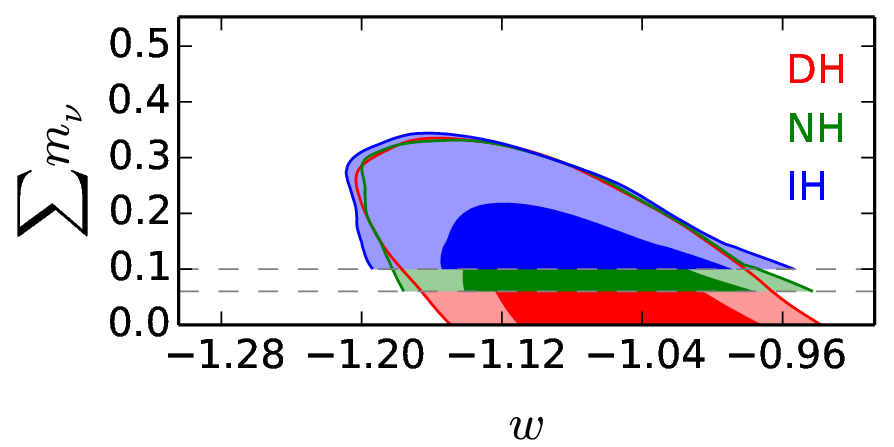}
\caption{The $68\%$ and $95\%$ CL marginalized contours of $\sum m_\nu$ and $w$ in the $w$CDM model by using the Planck TT,TE,EE + lowP + BAO + JLA + $H_0$ + Lensing data, in the case of considering neutrino mass hierarchies. }
\label{fig:sai_summnu_w_2D}
\end{figure}

\begin{figure}[htbp]
\centering
\includegraphics[width=3.4in]{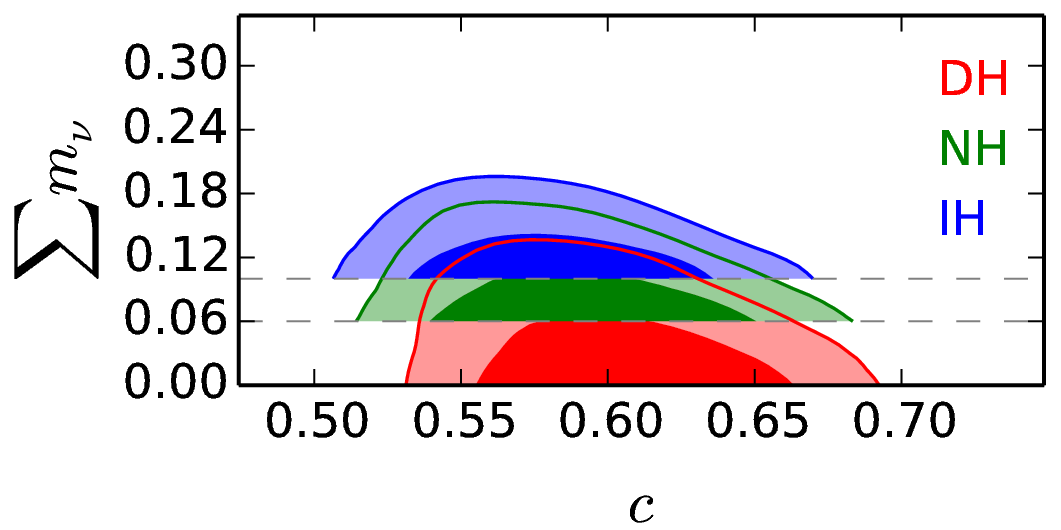}
\caption{The $68\%$ and $95\%$ CL marginalized contours of $\sum m_\nu$ and $c$ in the HDE model by using the Planck TT,TE,EE + lowP + BAO + JLA + $H_0$ + Lensing data, in the case of considering neutrino mass hierarchies.}
\label{fig:sai_summnu_c_2D}
\end{figure}

The dynamical dark energy models can significantly influence the measurement of the neutrino total mass $\sum m_\nu$.
The dependence of $\sum m_\nu$ upon the parameters of dynamical dark energy models is plotted in Figs.~\ref{fig:sai_summnu_w_2D} and \ref{fig:sai_summnu_c_2D}.
Both figures are obtained by using the Planck TT,TE,EE+lowP+BAO+JLA+$H_0$+Lensing data.

For the $\nu_\textrm{DH}w\textrm{CDM}$ model, from Table~\ref{tab:wcdm}, the constraint results of $\sum m_\nu$ and $w$ are $\sum m_\nu<0.273~\textrm{eV}$ and $w=-1.073\pm0.086$ without low-redshift data, and $\sum m_\nu<0.268~\textrm{eV}$ and $w=-1.080\pm0.050$ after considering low-redshift data. Obviously, including low-redshift data gives slightly tighter bounds, since it can tighten the constraint on the dark energy parameter $w$. Moreover, $\sum m_\nu$ is anti-correlated with the dark energy parameter $w$ in Fig.~\ref{fig:sai_summnu_w_2D}. After taking the neutrino mass splitting effect into account, the estimation of $w$ is driven to a lower value while the upper limit on $\sum m_\nu$ is driven to a higher value.

In HDE, since the upper bound of the neutrino total mass in DH model almost hit the lower bound of IH model, it's more appropriate to  consider neutrino mass splitting effect in the cosmological fit. From Table~\ref{tab:hde}, for $\nu_\textrm{NH}\textrm{HDE}$ model $\sum m_\nu<0.196~\textrm{eV}$ and $c=0.501\pm0.049$ without low-redshift data, and $\sum m_\nu<0.145~\textrm{eV}$ and $c=0.590\pm0.030$ with low-redshift data. For $\nu_\textrm{IH}\textrm{HDE}$ model $\sum m_\nu<0.223~\textrm{eV}$ and $c=0.488\pm0.047$ without low-redshift data, and $\sum m_\nu<0.173~\textrm{eV}$ and $c=0.581\pm0.029$ with low-redshift data. The neutrino mass $\sum m_\nu$ is anti-correlated with the dark energy parameter c in Fig.~\ref{fig:sai_summnu_c_2D}. Compared to DH model, the neutrino mass splitting effect lowers the estimation of $c$ while rises the upper limit on $\sum m_\nu$.

\begin{figure}[htbp]
\centering
\includegraphics[width=2.5in]{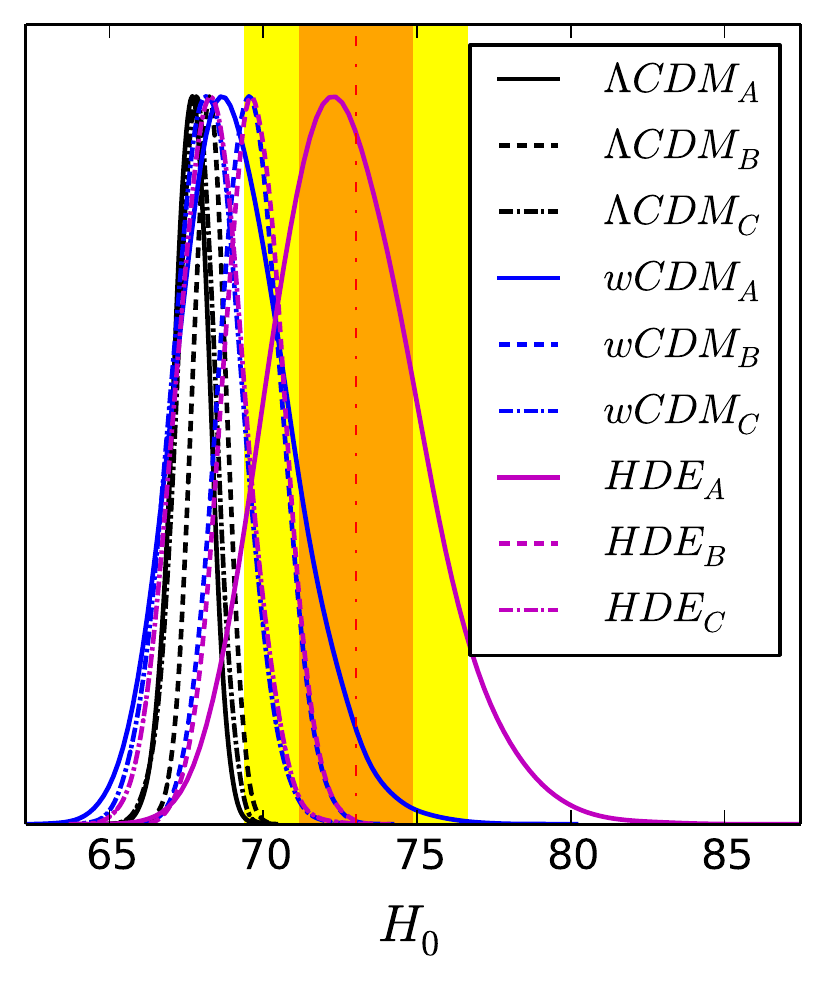}
\caption{The posterior probability distributions of $H_0$ in three dark energy models with three degenerate neutrinos. The red dot-dashed vertical line denotes the central value of the local $H_0$ measurement, and the orange (yellow) shaded area denotes the $1\sigma$ ($2\sigma$) uncertainty. Here $A$, $B$ and $C$ denote Planck TT,TE,EE+lowP+BAO, Planck TT,TE,EE+lowP+BAO+JLA+$H_0$+Lensing, and Planck TT,TE,EE+lowP+BAO+JLA+Lensing, respectively. }
\label{fig:sai_h0_1d}
\end{figure}

In our study, we have used the recent local measurement of $H_0$, namely, $H_0=73.02\pm1.79~\rm{km~s^{-1}~Mpc^{-1}}$ at $1\sigma$ CL \cite{Riess:2016jrr}. This measurement is in tension with the $\Lambda$CDM prediction based on Planck CMB observation. Figure~\ref{fig:sai_h0_1d} shows the posterior distributions of $H_0$ in three dark energy models with three degenerate neutrinos. The red dot-dashed vertical line denotes the central value of the local $H_0$ measurement, and the orange (yellow) shaded area denotes the $1\sigma$ ($2\sigma$) uncertainty. Here the subscript $A$, $B$ and $C$ denote Planck TT,TE,EE+lowP+BAO, Planck TT,TE,EE+lowP+BAO+JLA+$H_0$+Lensing and Planck TT,TE,EE+lowP+BAO+JLA+Lensing (excluding $H_0$ prior) data combinations, respectively. Our study shows that the $H_0$ tension still remains for $\Lambda$CDM and $w$CDM by using all the data combinations. For the $\Lambda$CDM model, the value of $H_0$ can be tightly constrained by Planck CMB data, due to the precise determination of the acoustic scale $\theta_\ast=r_s/D_A$. By contrast, this is not true for dynamical dark energy \cite{Li:2013dha}. From Fig.~\ref{fig:sai_h0_1d}, we find that $H_0$ is not well constrained by Planck CMB data in the dynamical dark energy models (see also \cite{Li:2013dha}). For the $w$CDM model, the uncertainty of $H_0$ becomes larger, and the $H_0$ tension is slightly alleviated, but not enough.  However, the best-fit result of $H_0$ in the HDE model by using Planck TT,TE,EE+lowP+BAO is well compatible with the local measurement of $H_0$.  After including the low-redshift data, however, the $H_0$ tension is recovered even though we have used the local measurement $H_0$ value as a prior. Our results on $H_0$ in dynamical dark energy models with massive neutrinos are consistent with those in \cite{Li:2013dha} where massive neutrinos are not considered. For comparison, we also show what the model fitting does when the $H_0$ prior is excluded in Figure~\ref{fig:sai_h0_1d}. For the $w$CDM model,  the best-fit $H_0$ value is changed to $68.2\pm1.1~\rm{km~s^{-1}~Mpc^{-1}}$, while for the HDE model, it is changed to $68.3\pm1.1~\rm{km~s^{-1}~Mpc^{-1}}$. It shows that the best-fit $H_0$ value is only different from that including $H_0$ prior by around 1$\sigma$. By careful check on our best-fit results, we find that the constraints on neutrino total mass  including $H_0$ prior are only slightly different from those excluding $H_0$ prior.

Actually, we have shown that replacing the $\Lambda$CDM model with a dynamical dark energy model is indeed helpful for relieving the tension between the local measurement and the Planck constraint results of $H_0$, but not enough. Considering the extra relativistic degrees of freedom, i.e., an additional parameter $N_{\rm eff}$, is more helpful for this problem (for a detailed discussion, see \cite{Zhang:2014dxk}). It has been shown in \cite{Zhang:2014dxk,Dvorkin:2014lea,Zhang:2014nta} that the involvement of massive sterile neutrinos in the cosmological model could simultaneously relieve almost all the tensions among the astrophysical observations, leading to a new cosmic concordance. Another study recently showed that the tension of $H_0$ can be resolved in an extended parameter space \cite{DiValentino:2016hlg}. We leave the further discussions on this issue to a future study.




\section{Conclusion}\label{conclusion}

The dynamical dark energy can significantly influence the constraints on the neutrino total mass $\sum m_{\nu}$. In this paper, we studied two typical dynamical dark energy models, namely, the $w$CDM model and the HDE model, by using two recent data combinations which are denoted by Planck TT,TE,EE+lowP+BAO and Planck TT,TE,EE+lowP+BAO+JLA+$H_0$+Lensing, respectively. By contrast to the $\Lambda$CDM model, the upper limit on $\sum m_{\nu}$ becomes much looser in the $w$CDM model while tighter in the HDE model. This is consistent with the previous study \cite{Zhang:2015uhk}. For all the dark energy models considered in this paper, the minimal $\chi^2$ is slightly smaller in the NH case than that in the IH case. Thus the NH models fit two data combinations better than the IH ones. Even so, the difference $\Delta \chi^2$ is still not significant enough to distinguish the neutrino mass hierarchy. Even worse, the DH models fit both data combinations best. In addition, we found that the local measurement of $H_0$ is compatible with the best-fit value of $H_0$ in the HDE model by using Planck TT,TE,EE+lowP+BAO. Once more low-redshift observations were added, however, the $H_0$ tension was found to be recovered.

For the $\nu_{\textrm{DH}}$HDE model, we have used the Planck TT,TE,EE+lowP+BAO+JLA+$H_0$+Lensing data to obtain the $95\%$ upper limit on the neutrino mass, i.e. $\sum m_{\nu}<0.105~\textrm{eV}$, which is comparable to the lower limit of $\sum m_{\nu}$ for three inverted hierarchical neutrinos. We have already stood on the verge to distinguish the neutrino mass hierarchy through cosmological observations. This constraint is more stringent than the previous study \cite{Zhang:2015uhk}. To our knowledge, this is perhaps the most stringent upper limit on the total mass of three degenerate neutrinos up to now. Thus the neutrino mass window could be much tighter if the accelerating expansion of the Universe is not driven by the cosmological constant. However, more observational data are needed to further study the neutrino sector. Future observations, for instance, BAO \cite{Zhao:2015gua,Font-Ribera:2013rwa}, CMB \cite{Calabrese:2014gwa,Benson:2014qhw,Matsumura:2013aja,Kogut:2011xw}, and galaxy shear surveys \cite{  Abell:2009aa,Laureijs:2011gra} in near future, might reach the sensitivity to determine the neutrino mass and to distinguish the mass hierarchy of three active neutrinos.


\acknowledgments
We acknowledge the use of CosmoMC.
DMX is supported by the National Natural Science Foundation of China (Grant No. 11505018) and the Chongqing Science and Technology Plan Project (Grant No. Cstc2015jvyj40031). XZ is supported by the Top-Notch Young Talents Program of China, the National Natural Science Foundation of China (Grants No. 11522540 and No. 11175042), and the Fundamental Research Funds for the Central Universities (Grants No. N140505002 and No. N140506002). This work is partially supported by a grant from the Research Grant Council of the Hong Kong Special Administrative Region, China (Project No. 14301214).



\end{document}